\documentclass[a4paper,11pt]{article}
\usepackage{jinstpub} 
\usepackage{lineno}
\usepackage{subcaption}
\usepackage{multirow}

\title{Measurements of proton hardness factors in silicon at energies between 10 MeV and 25 MeV}

\author[a, 1]{E. Liu, \note{Corresponding author.}}
\author[a, 1]{A. Chisholm, }
\author[a]{L. Gonella, }
\author[a]{A. Hunter }
\affiliation[a]{School of Physics and Astronomy,\\University of Birmingham,\\
Birmingham, B15 2TT, United Kingdom}
\emailAdd{ehl857@student.bham.ac.uk, a.s.chisholm@bham.ac.uk}

\abstract{Silicon detector technologies are often employed for high energy particle physics applications due to their excellent radiation hardness. Radiation damage in the form of bulk or substrate damage is dependent on the incident particle species and energy. The fluence is therefore often quoted as the 1 MeV neutron equivalent fluence, with hardness factors being the scaling quantity. The hardness factor can be determined by analysing the change in the leakage current of silicon pad diodes post-irradiation. Using the MC40 cyclotron facility at the University of Birmingham, n-in-p FZ silicon pad diodes were irradiated to several fluence points at three different proton beam energies. The hardness factors acquired were 3.07 $\pm$ 0.37 for 10.5 MeV protons, 2.73 $\pm$ 0.27 for 16.4 MeV protons and 2.19 $\pm$ 0.22 for 24.3 MeV protons. The value for 16.4 MeV protons agrees with the theoretical predictions, whereas the values for 10.5 MeV and 24.3 MeV are lower than predicted by the same calculations. }


\arxivnumber{2408.12410} 

\begin{document}
\maketitle
\flushbottom

\section{Introduction}
\label{sec:intro}

Silicon detector technologies are employed in high energy particle physics applications due to their excellent radiation hardness among other desirable characteristics. A potential effect of incident radiation on a silicon sensor is bulk (substrate) damage, associated with ion displacements caused by non-ionising energy losses (NIEL) in the silicon lattice. Silicon atoms which are displaced from their lattice sites lead to defects in the lattice. One consequence is an increase in the detector leakage current. Bulk damage is dependant on many factors including the incoming particle species and energy. To facilitate comparisons when quantifying bulk damage, fluences are conventionally scaled to the equivalent fluence of 1 MeV neutrons under the assumption of the NIEL scaling hypothesis \cite{NIELMention}. The parameter used to scale the fluence is called the hardness factor, denoted $\kappa$.


This article will present the measurements of the proton hardness factor in silicon for three different proton energies delivered using the MC40 cyclotron facility at the University of Birmingham. A previous measurement at this facility for 24 MeV protons determined a hardness factor of 2.1 $\pm$ 0.5 \cite{Allport:2019kvs}. For this earlier measurement, commercial BPW34 silicon photodiodes were irradiated to several fluences before performing current-voltage (IV) measurements to determine the change in leakage current associated with displacement damage. This study extends the measurements to lower proton energies, together with the aim of improving the precision on the hardness factor measurements. For this measurement, n-in-p float zone (FZ) silicon diodes were chosen, as this is the typical technology used in many current and future particle physics experiments \cite{ATLASITkStripTDR, ATLAS18SensorFZ, CMSTrackerTDR, CMSStripFZ}.

\section{MC40 Cyclotron Facility}
\label{sec:MC40}
The MC40 cyclotron at the University of Birmingham is primarily used for the production of radioisotopes for medical applications. It is also utilised to perform irradiations for many research programs, including radiation damage studies in silicon sensors. The facility is able to provide proton beams with an energy from 2 MeV to 38 MeV at currents up to a few $\muup$A.

\begin{figure}[htbp]
\centering
\includegraphics[width=.85\textwidth, height=0.31\textheight]{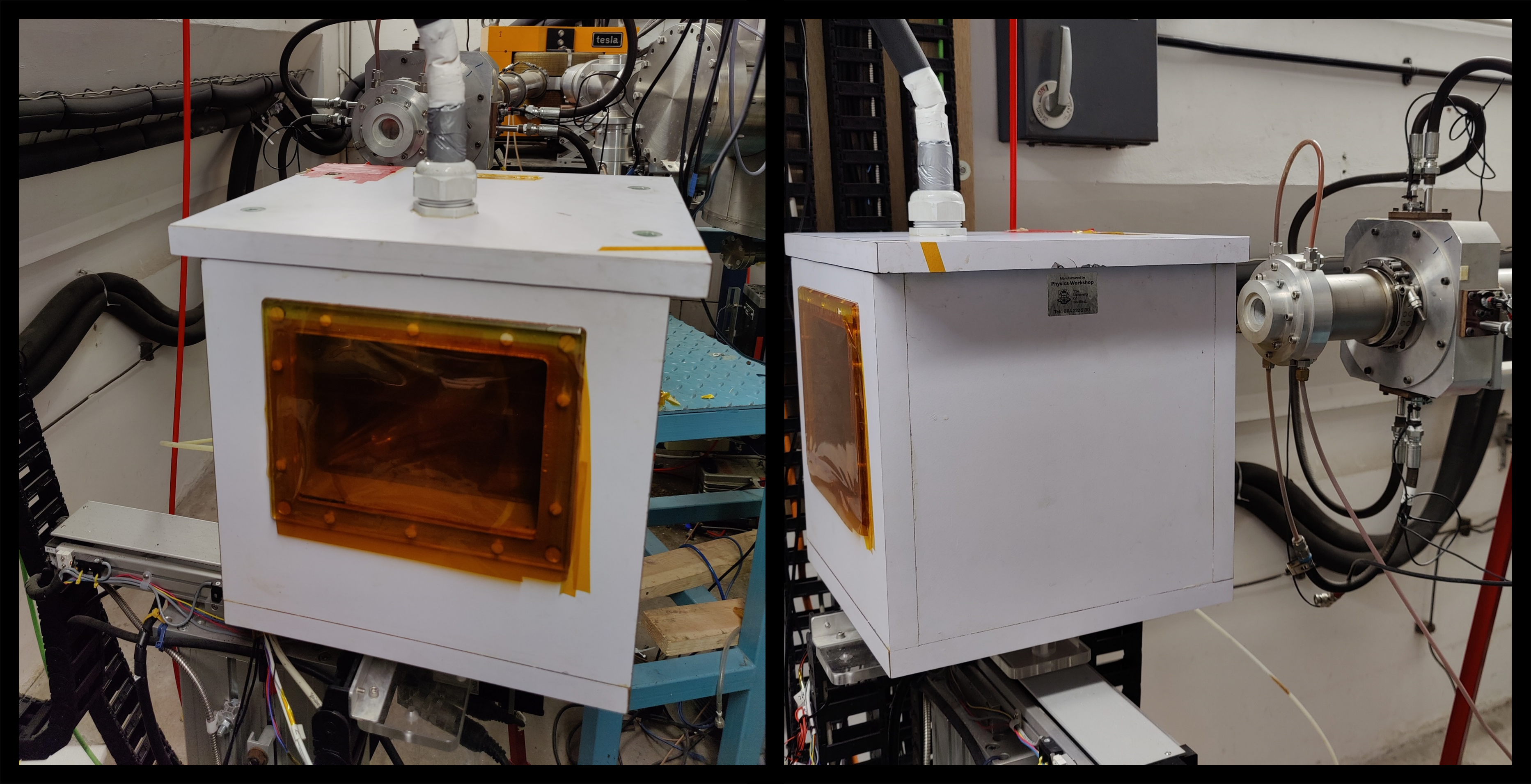}
\caption{Two views of the cyclotron beamline used for silicon sensor irradiations, showing the beam exit and the insulated box.\label{fig:DetectorBeamline}}
\end{figure}

\begin{figure}[htbp]
\centering
\includegraphics[width=.99\textwidth, height=.35\textheight]{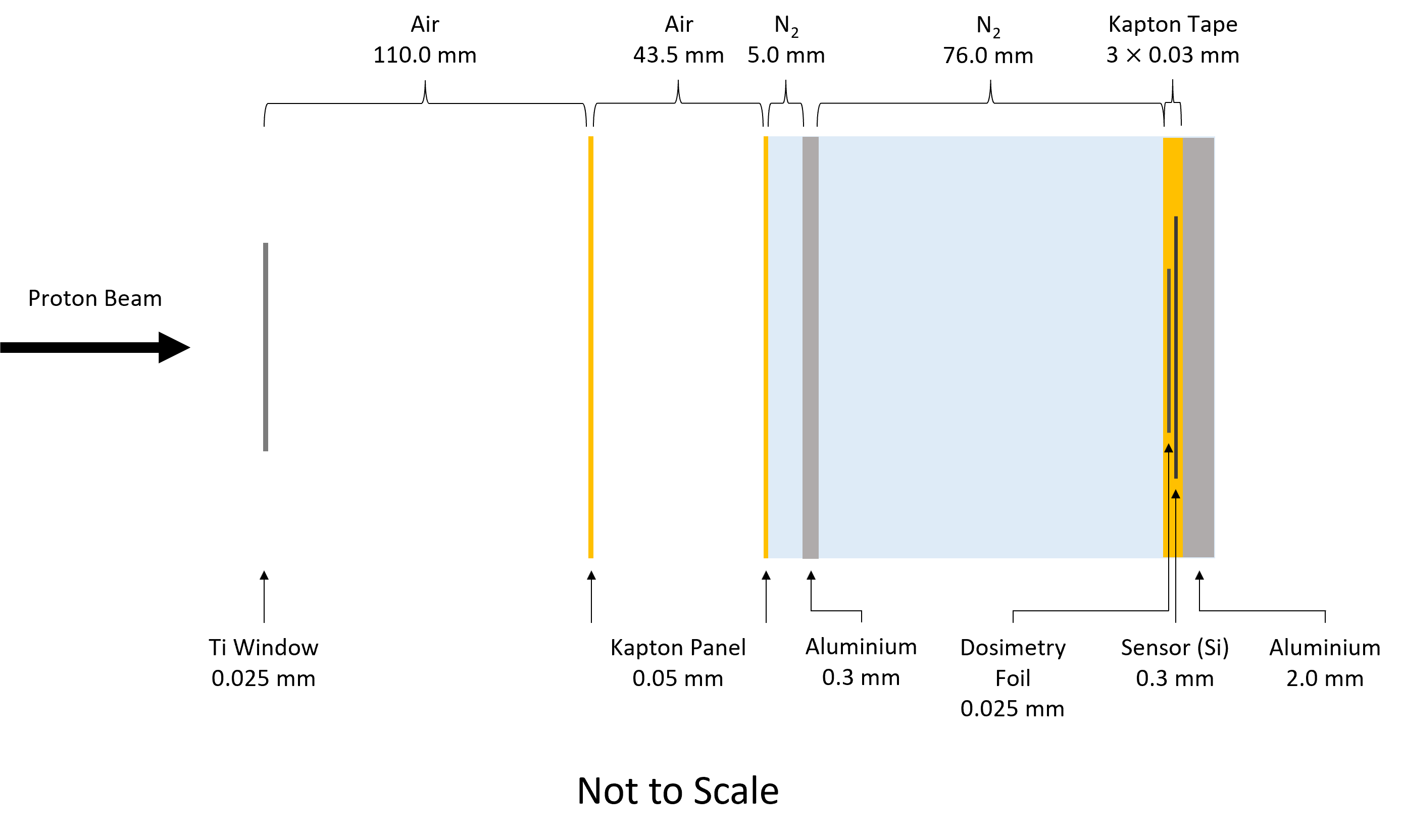}
\caption{A cross-section of the irradiation setup showing the sensor sample, dosimetry foil and the surrounding media.\label{fig:BeamlineCS}}
\end{figure}

Figure~\ref{fig:DetectorBeamline} shows the irradiation setup used for silicon sensor irradiations. The material traversed by the beam can be seen in Figure~\ref{fig:BeamlineCS}, showing the material thicknesses relevant for this investigation. The proton beam is collimated into a roughly $10 \times 10$ mm$^{2}$ square area with approximately uniform intensity, emerging from a thin titanium window placed at the end of the beamline. The samples for irradiation are mounted on an aluminium plate, itself mounted inside an insulated box. An internal temperature of -27$^{\circ}$C is maintained by liquid nitrogen evaporative cooling to ensure that thermal annealing effects are suppressed when irradiating samples at high dose rates. The nitrogen atmosphere also maintains a relative humidity below 10\%. Samples are secured on the aluminium plate with kapton tape. In front of each sample is a 25 $\muup$m thick piece of nickel or titanium foil, used to determine the proton fluence delivered to that sample after irradiation by radioisotope activation analysis. Inside the box, a 300 $\muup$m sheet of aluminium is placed to remove any potential low energy components of the beam. The box is mounted on an XY-motorised stage controlled using LabView. A grid-scanning pattern is employed to deliver a uniform fluence to all samples.

\begin{figure}[htbp]
\centering
\includegraphics[width=.9\textwidth]{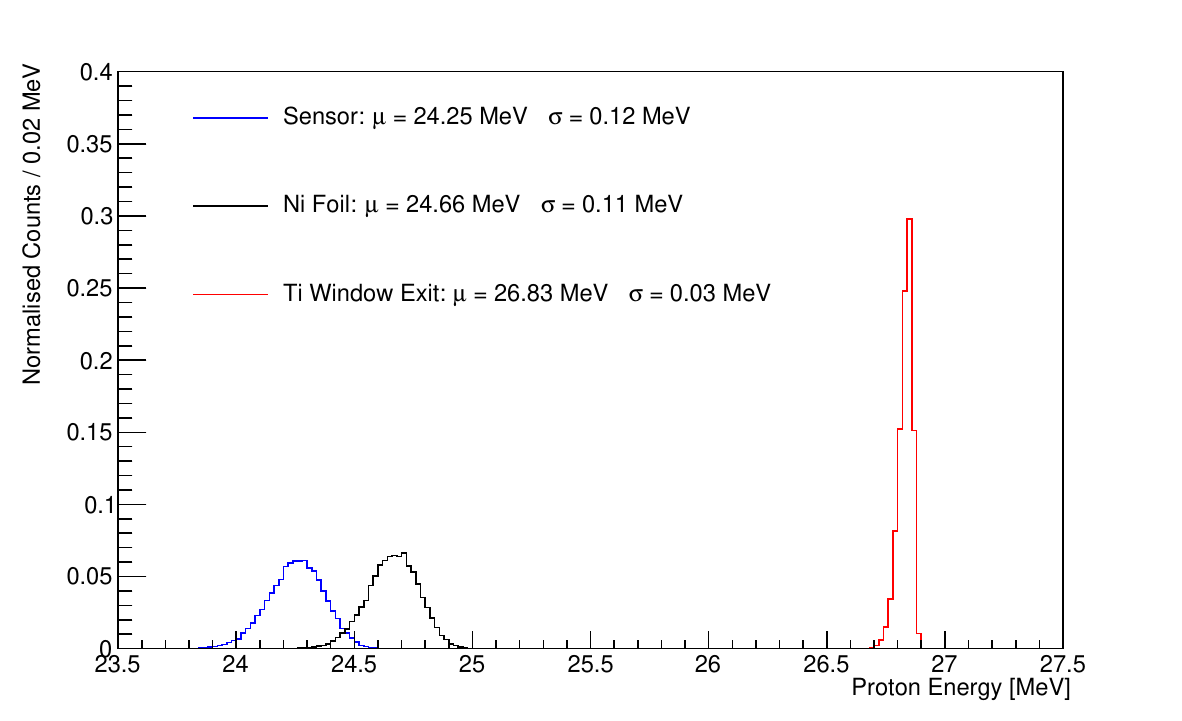}
\caption{The beam energy profile incident on the Ni foil and sensor from a 27 MeV cyclotron beam energy. The mean ($\muup$) and standard deviation ($\sigma$) of the distributions are also shown.\label{fig:27MeVBeamEnergy}}
\end{figure}

A Geant4 \cite{GEANT4:2002zbu} simulation of the beamline was developed to estimate the evolution of the proton beam energy throughout the setup. Figure~\ref{fig:27MeVBeamEnergy} shows the beam energy directly emerging from the vacuum, incident on both the nickel foil and the sensor for a 27 MeV beam energy. The beam energies incident on the foil and sensor are important for fluence measurement and understanding the proton energy range relevant for the hardness factor measurement, respectively. Similar simulations were performed for the 20 MeV and 15 MeV beam energies. Table~\ref{tab:G4ProtonEnergy} summarises the proton energy at the sensor for the beam energies relevant to this study.

\begin{table}[htbp]
\centering
\caption{The Geant4 simulation of the proton energies at the dosimetry foil and the sensor for a given beam energy.\label{tab:G4ProtonEnergy}}
\smallskip
\begin{tabular}{|c|c|c|}
\hline
Proton Beam Energy [MeV] & Energy at Dosimetry Foil [MeV] & Energy at Sensor [MeV]\\
\hline
27 & 24.7 $\pm$ 0.1 & 24.3 $\pm$ 0.1\\
20 & 17.0 $\pm$ 0.1 & 16.4 $\pm$ 0.1\\
15 & 11.0 $\pm$ 0.1 & 10.5 $\pm$ 0.1\\
\hline
\end{tabular}
\end{table}

\section{Proton Fluence Determination}
\label{sec:dosimetry}

The proton fluence of the irradiation is determined offline with foil activation analysis. Natural nickel foil was chosen for the 27 MeV and 20 MeV beam energy runs and natural titanium foil for the 15 MeV energy run. This choice optimises the activation cross-section for the target monitor reaction in the relevant energy range \cite{HERMANNE2018338}. The two monitor reactions $^{\text{nat}}$Ni($p, x$)$^{57}$Ni and $^{\text{nat}}$Ti($p, x$)$^{48}$V were chosen for the high relative gamma-decay branching ratios of the nuclides produced. The activity of the foils was measured using a high-purity germanium (HPGe) detector. The proton fluence is then calculated using the activity, the cross-section of the beam monitor reaction and the mass of the measurement foils \cite{STRIJCKMANS200510, MASTREN20191}.

\section{Measurement Procedure}
\label{sec:measurements}







The devices selected for this study are monitor diodes of area approximately $8\times8$ mm$^{2}$ (MD8) on ATLAS ITk strip sensor quality assurance samples \cite{Ullan:2020juz}. These are n-in-p silicon diodes surrounded by a guard ring, with a bulk thickness of approximately 295 $\muup$m and an active area of 0.5095 cm$^{2}$.

\begin{figure}[htbp]
\centering
\includegraphics[width=.9\textwidth]{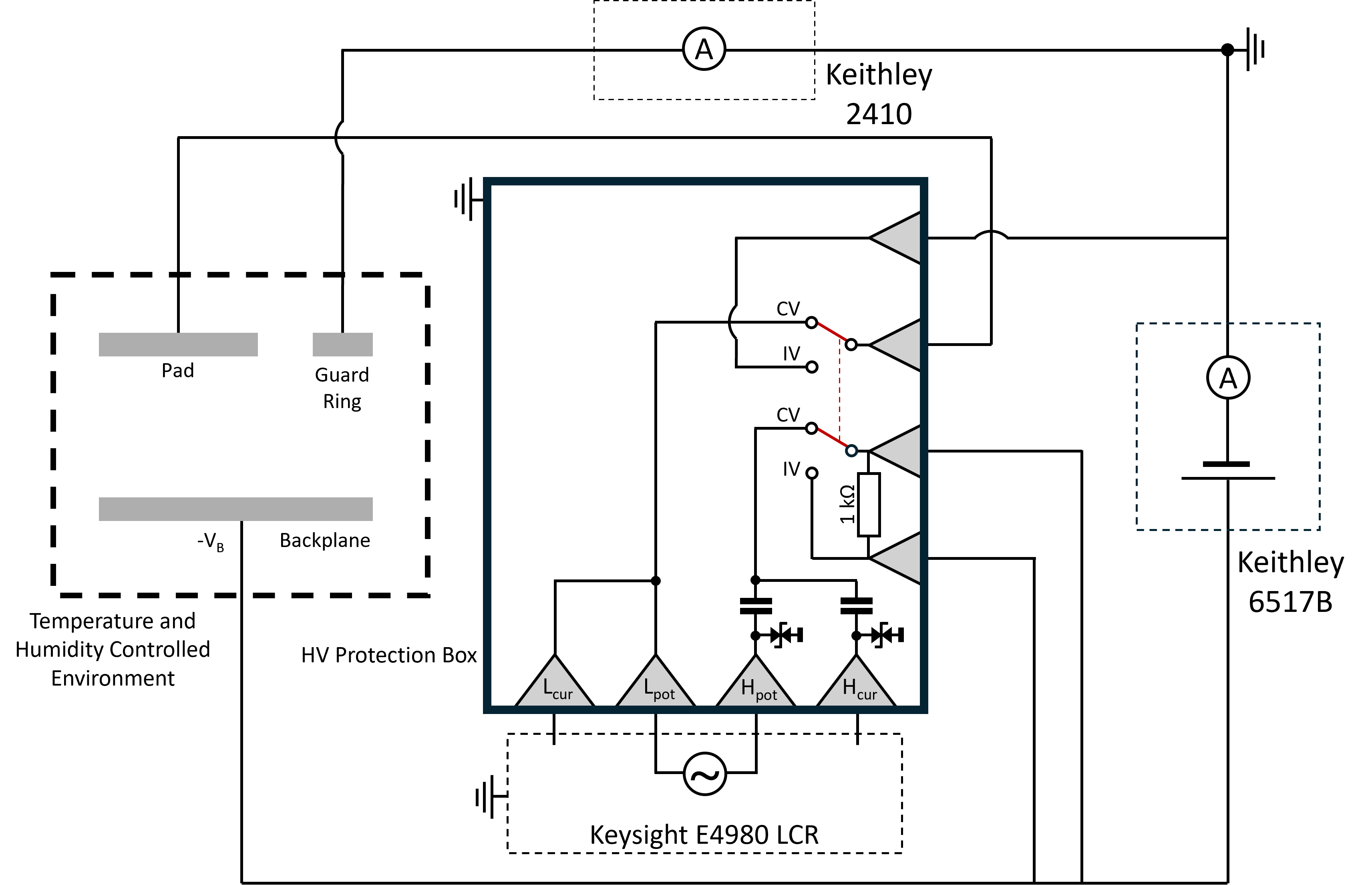}
\caption{The circuit diagram for the IV and CV measurements. The grey triangles denote connections between the sealed HV protection box and the external circuit via BNC cables. A double-pole switch allows the circuit to be changed between IV and CV measurement modes. The temperature and humidity controlled environment represents the climate chamber and nitrogen flushing.\label{fig:CircuitDiagram}}
\end{figure}

In order to facilitate IV and capacitance-voltage (CV) measurements, sensors are attached to a custom-made PCB using silver tape before wire bonding. The PCB under test is housed in a testing stage within a climate chamber, used to regulate the temperature. The testing stage is flushed with dry nitrogen to maintain a relative humidity below 10\%. Irradiated samples first undergo thermal annealing at 60$^{\circ}$C for 80 minutes. Measurements are then performed at -20$^{\circ}$C in order to regulate the leakage current and avoid sensor self-heating. Figure~\ref{fig:CircuitDiagram} is a circuit diagram showing the instrument connections used to perform the IV and CV measurements.

\subsection{Leakage Current Measurements}

\begin{figure}[htbp]
\centering
\includegraphics[width=.8\textwidth]{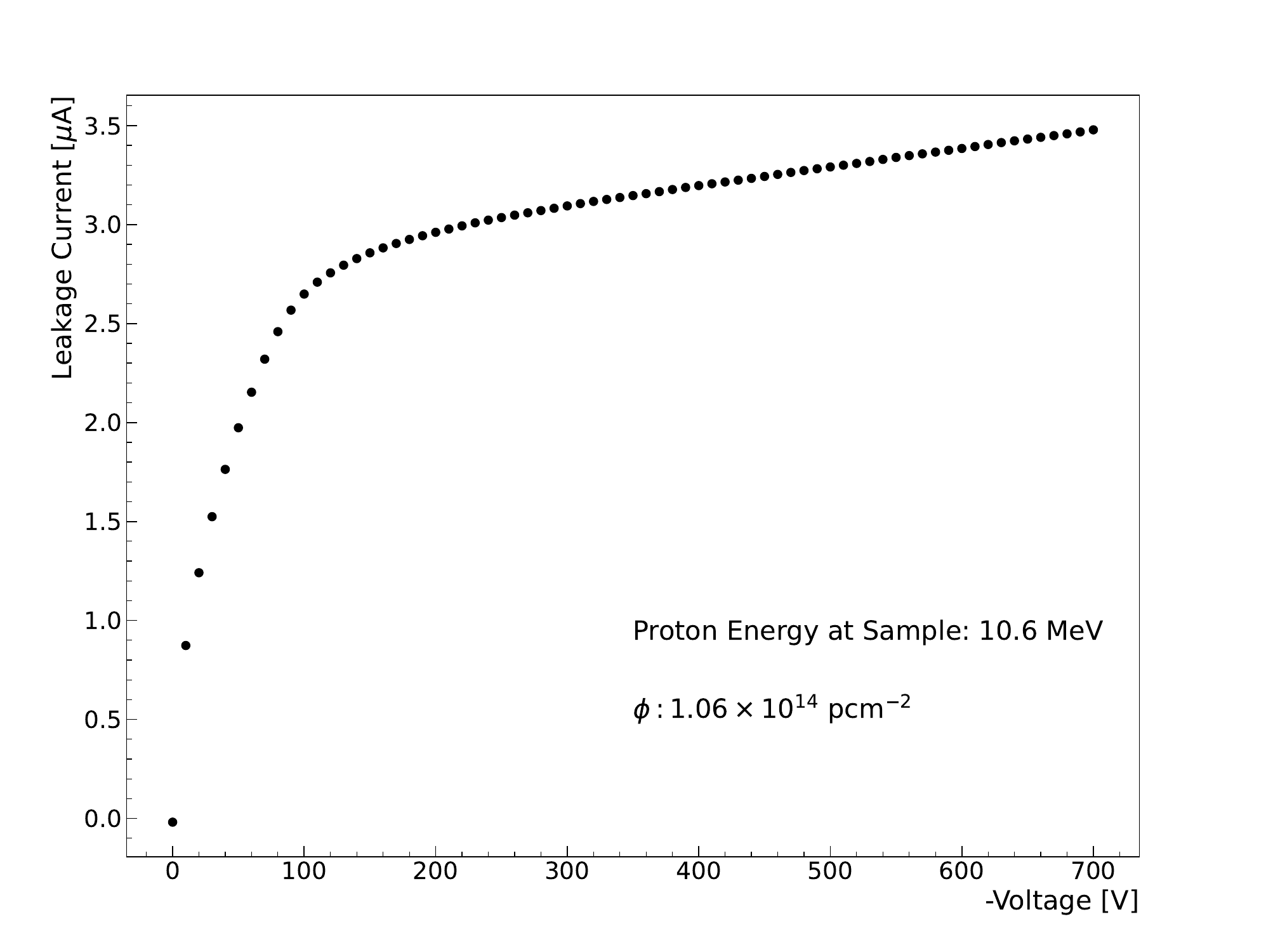}
\caption{The IV curve for a sample irradiated to a proton fluence of 1.06 $\times$ 10$^{14}$ cm$^{-2}$ with 15 MeV protons. \label{fig:ExampleIV}}
\end{figure}

The leakage current is measured with a Keithley 6517B electrometer which supplies the high voltage bias to the backplane of the sensor in addition to measuring the leakage current through the diode pad. The guard ring is connected to a Keithley 2410 sourcemeter which is common grounded to the pad ground. A voltage sweep is performed up to -700 V in 10 V increments, remaining well below the breakdown voltage of these structures. Figure~\ref{fig:ExampleIV} shows an example MD8 IV curve for a sensor irradiated to a proton fluence around $1 \times 10^{14}$ cm$^{-2}$ with 15 MeV protons. 

\label{sec:IVMeas}
\subsection{Full Depletion Voltage Measurements}
\begin{figure}[htbp]
\centering
\includegraphics[width=.8\textwidth]{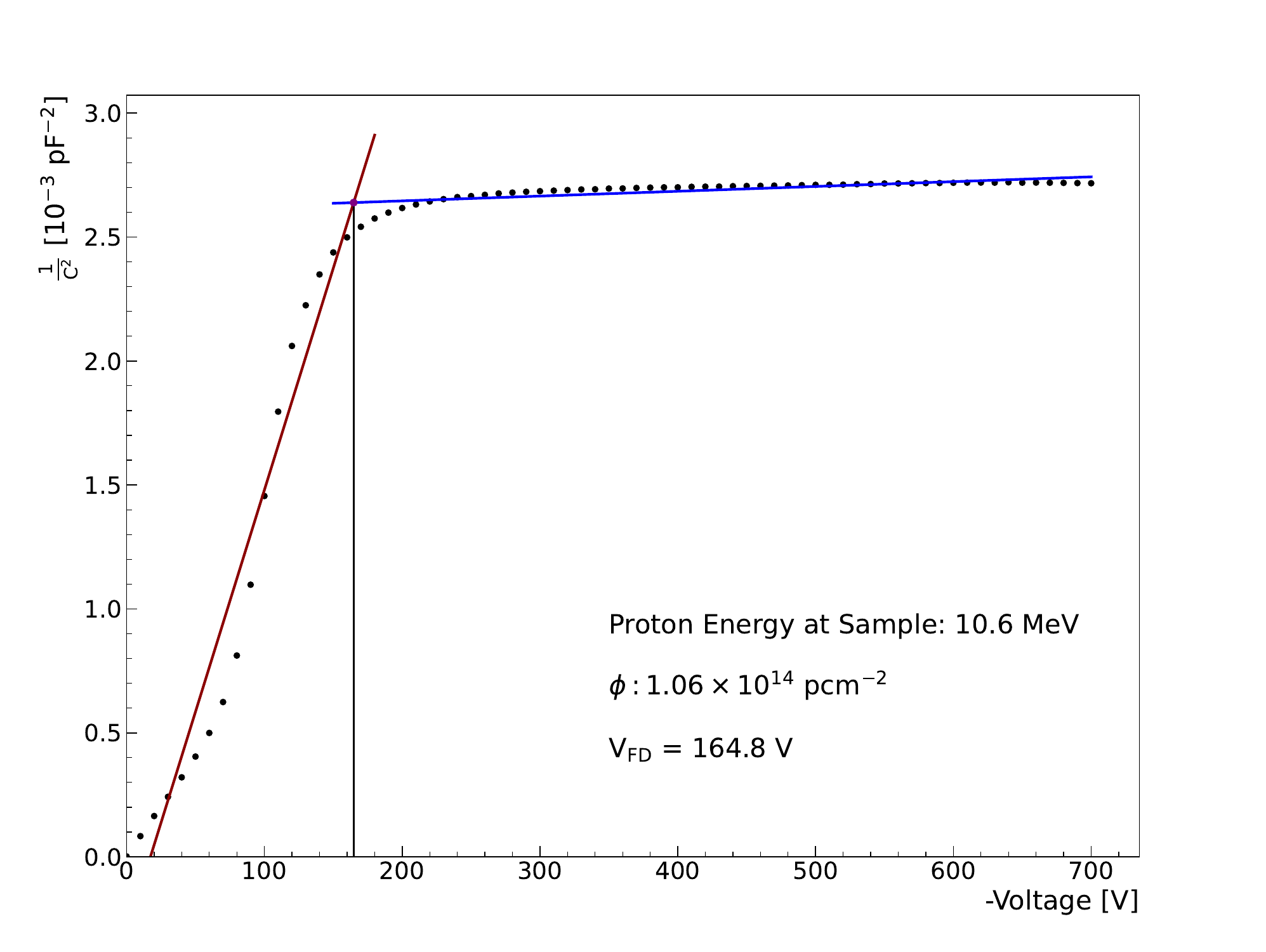}
\caption{The $\text{C}^{-2}$ vs V curve for a sample irradiated to a proton fluence of 1.06 $\times$ 10$^{14}$ cm$^{-2}$ with 15 MeV protons. The red and blue lines are fits to the linear and constant regions respectively.\label{fig:ExampleCVforVFD}}
\end{figure}

For silicon pn-junction diodes, the inverse-square of the capacitance is expected to increase linearly with voltage as the depletion region grows. At and beyond full depletion, the capacitance becomes constant. From a plot of $\text{C}^{-2}$ versus voltage, two linear fits are drawn on the linear and constant regions, with their intersection providing the estimate of the full depletion voltage (V$_{\mathrm{FD}}$) for the sensor. An example curve is given in Figure~\ref{fig:ExampleCVforVFD}, for a sample irradiated to a proton fluence around 1 $\times$ 10$^{14}$ cm$^{-2}$ with 15 MeV protons. The extracted V$_{\text{FD}}$ is 164.8 V. Across all samples for the three different proton energies, the V$_{\text{FD}}$ does not vary substantially. The V$_{\mathrm{FD}}$ is important when extracting the leakage current from the IV curves for the hardness factor determination as this should be after the bulk is fully depleted and at a consistent value across all samples.

The capacitance measurements utilise the same Keithley 6517B and 2410 devices to supply the high bias voltage to the sensor and to ground the guard ring, respectively. The capacitance is measured using a Keysight Agilent E4980A LCR meter. A high voltage protection circuit is used to isolate the 6517B bias from the LCR inputs. An open correction is performed prior to each measurement. A frequency of 1 kHz and a 100 mV amplitude was selected for the AC measurement signal.

\label{sec:CVmeas}

\section{Hardness Factor Determination}
\label{sec:HFCalc}

The leakage current in silicon diodes is expected to increase proportionally with fluence after irradiation from the radiation induced defects primarily introduced in the silicon bulk. The change in the leakage current as a function of fluence for a given sensor is governed by Eq.~\eqref{eqn:LeakageCurrent}:
\begin{equation}
    \Delta I = \alpha A w \phi, \label{eqn:LeakageCurrent}
\end{equation}
where $A$ is the active area of the sensor, $w$ is the depletion width and $\phi$ is the proton fluence \cite{MollAlphaNeq}. The parameter $\alpha$ is known as the current-related damage factor and links the change in leakage current to the fluence. From the NIEL scaling hypothesis, the hardness factor may be expressed as the ratio of the current-related damage factor for the protons relative to that for 1 MeV neutrons, given by Eq.~\eqref{eqn:MeasHF}:
\begin{equation}
    \kappa = \frac{\alpha}{\alpha_{\text{n}_{\text{eq}}}}, \label{eqn:MeasHF}
\end{equation}
where $\alpha_{\text{n}_{\text{eq}}}$ is the current-related damage factor for 1 MeV neutrons. The value of $\alpha_{\text{n}_{\text{eq}}}$ in silicon is found to be (3.99 $\pm$ 0.03) $\times$ 10$^{-17}$ A\,cm$^{-1}$ after annealing for 60$^{\circ}$C for 80 minutes and for currents measured at 20$^{\circ}$C \cite{MollAlphaNeq}.

The leakage current for the irradiated IV measurements taken at -20$^{\circ}$C are scaled to 20$^{\circ}$C using Eq.~\eqref{eqn:CurrentScale}:
\begin{equation}
    \text{I}(\text{T}_\text{R}) = \text{I}(\text{T})\cdot\left(\frac{\text{T}_\text{R}}{\text{T}}\right)^2\cdot e^{-\frac{\text{E}_\text{a}}{2\text{k}_\text{B}}\left(\frac{1}{\text{T}_\text{R}} - \frac{1}{\text{T}}\right)}, \label{eqn:CurrentScale}
\end{equation}
where T$_{\text{R}}$ is the target temperature of 20$^{\circ}$C and E$_{\text{a}}$ is the activation energy of silicon \cite{MollAlphaNeq}. A value of 1.21 eV was chosen for E$_{\text{a}}$, motivated by the temperature range used in this study and the study described in Ref. \cite{A_Chilingarov_2013}. All currents were scaled to the same temperature of 20$^{\circ}$C.

The leakage current must be extracted from the IV curve at a bias voltage greater than the full depletion voltage. From the measurements described in Section~\ref{sec:CVmeas}, the extraction voltage is chosen at -500 V for all tested sensor samples as it is expected that all sensors are fully depleted at this point.

To extract the current-related damage factor, a least squares fit is performed on the data in accordance to Eq.~\eqref{eqn:LeakageCurrent}. The $\alpha$ is a free parameter, whereas $A$ and $w$ are constants of the fit. The uncertainty in $\alpha$ associated with the finite precision of the current measurements is assessed directly with the least squares fit. However, the corresponding uncertainty in $\alpha$ associated with the fluence measurement uncertainties requires a dedicated approach. As described in Section~\ref{sec:results}, the uncertainty in the fluence of each sample consists of three main components, namely a statistical component associated with the measured number of counts on the HPGe detector, the uncertainties associated with the monitor reaction cross-sections and the efficiency of the HPGe detector. The correlation behaviour of these uncertainties between different samples varies from being entirely correlated (e.g. detector efficiency) to entirely uncorrelated (e.g. counting uncertainty). As such, the overall correlations in the fluence uncertainties for samples used to determine each hardness factor are non-trivial and a Monte Carlo approach \cite{BootstrapMC} is used to accurately propagate their effects to estimate the corresponding uncertainty on $\alpha$.



\section{Results}
\label{sec:results}

\begin{figure}[htbp]
\centering
\subfloat[]{\label{fig:27MeVHF}\includegraphics[width=.65\linewidth, height=0.281\textheight]{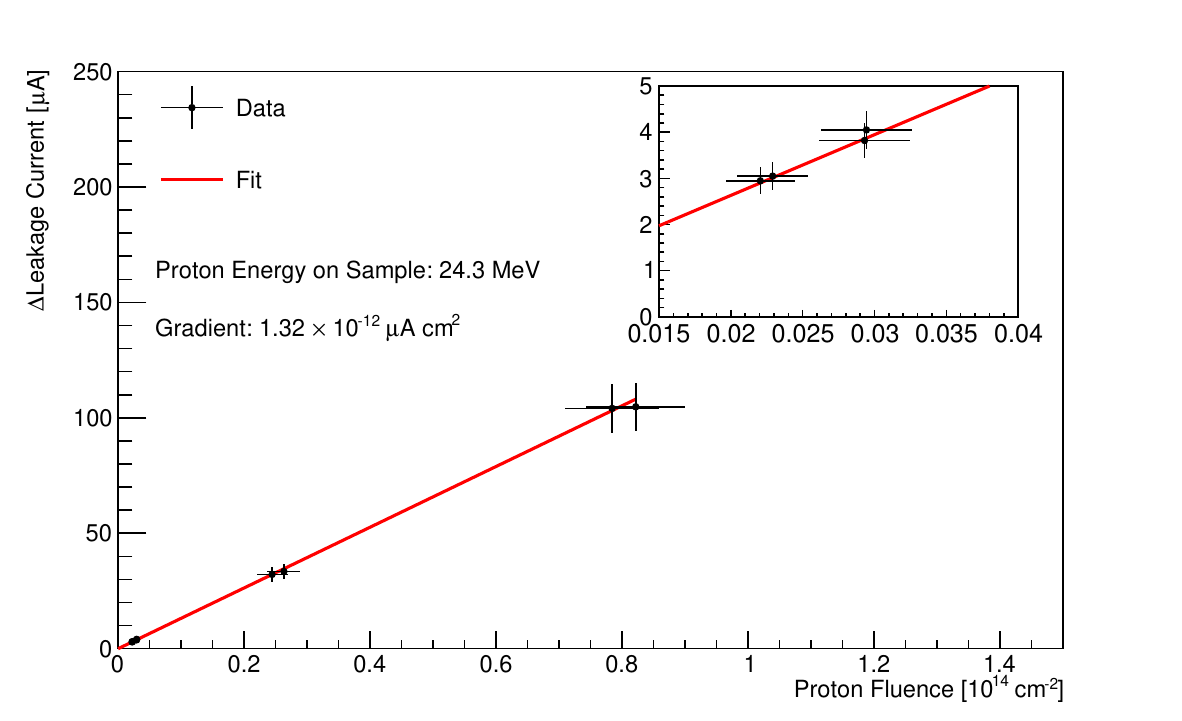}}\par
\subfloat[]{\label{fig:20MeVHF}\includegraphics[width=.65\linewidth, height=0.281\textheight]{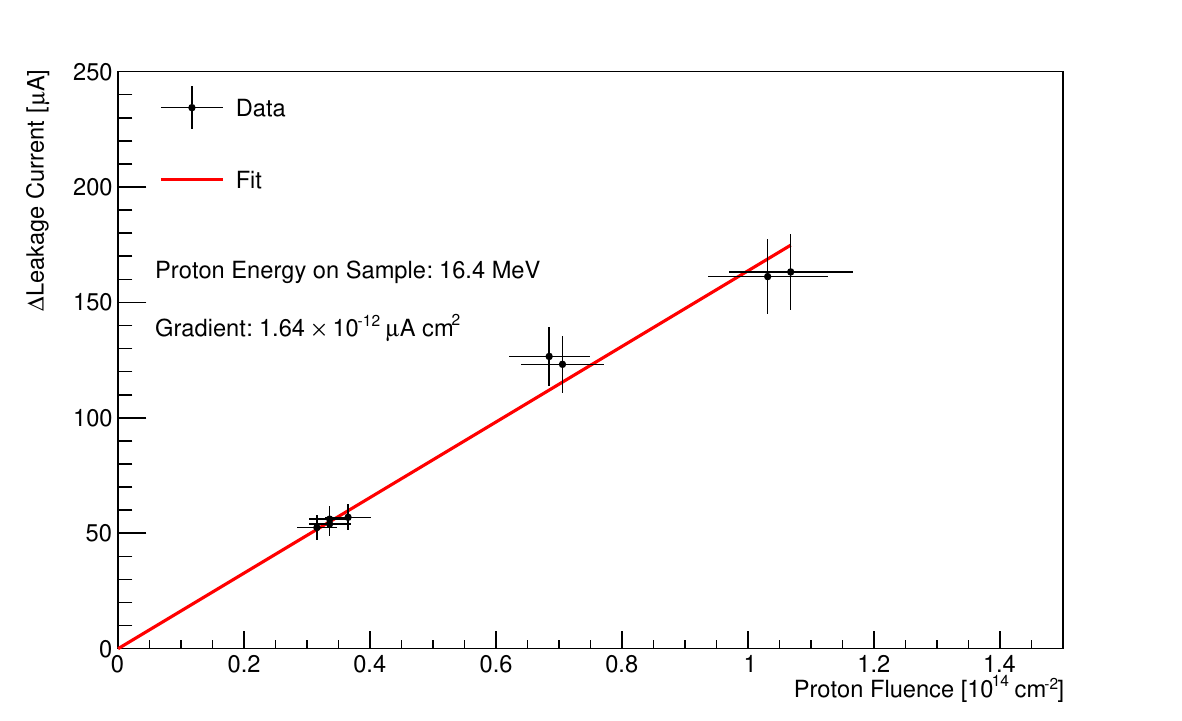}}\par 
\subfloat[]{\label{fig:15MeVHF}\includegraphics[width=.65\linewidth, height=0.281\textheight]{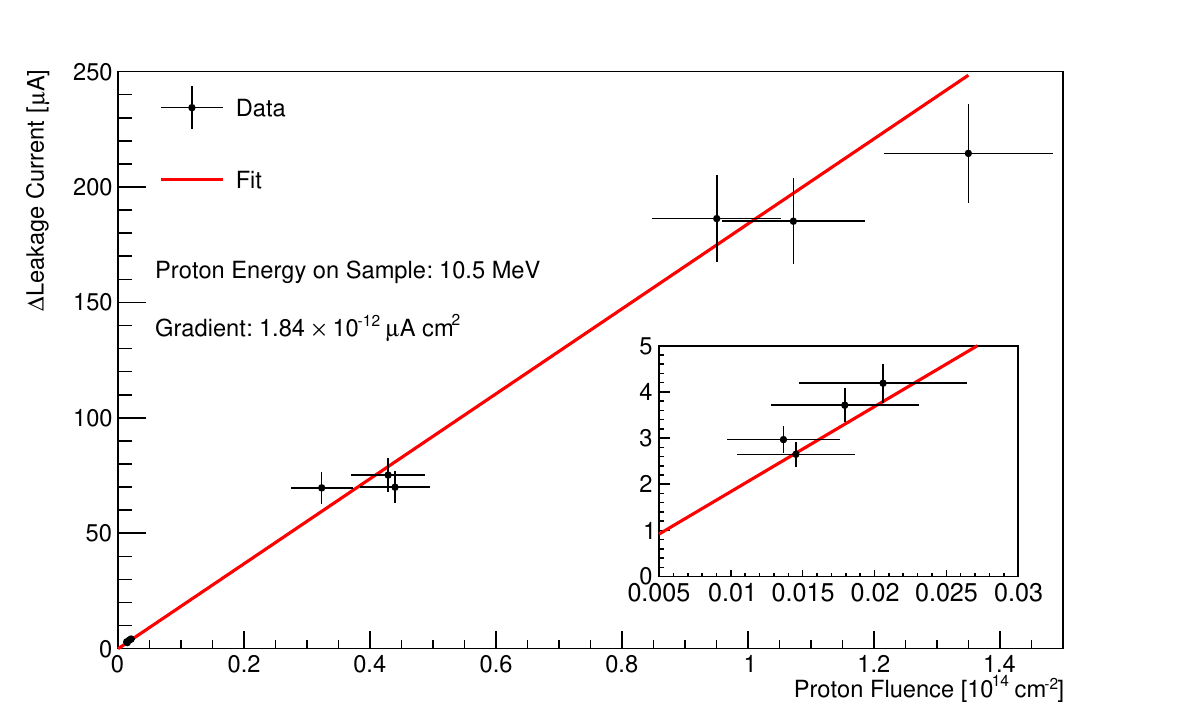}}
\caption{Measured change in leakage currents as a function of proton fluence for (\subref{fig:27MeVHF}) 24.3 MeV, (\subref{fig:20MeVHF}) 16.4 MeV and (\subref{fig:15MeVHF}) 10.5 MeV protons on the monitor diode.}
\label{fig:HFMeasPlots}
\end{figure}

The change in the leakage current through the diode as a function of proton fluence for proton energies of 24.3 MeV, 16.4 MeV and 10.5 MeV are shown in Figures~\ref{fig:27MeVHF},~\ref{fig:20MeVHF} and~\ref{fig:15MeVHF}, respectively, along with the least squares fit. In all three cases, the fit function was forced to go through the origin as per Eq.~\eqref{eqn:LeakageCurrent}. The leakage current of the samples was only measured post-irradiation, however, many measurements on non-irradiated samples from prior studies have shown that the typical pre-irradiated leakage currents are many orders of magnitude lower. Therefore, to a good approximation, the current measured post-irradiation can be expressed as the change due to the fluence. Table~\ref{tab:HFResults} summarises the current-related damage factors and their corresponding measured hardness factors for different proton energies on the silicon diode.


\begin{table}[htbp]
\centering
\caption{The measured current-related damage factors and the corresponding hardness factors for different proton energies, measured in silicon.\label{tab:HFResults}}
\smallskip
\begin{tabular}{|c|c|c|}
\hline
Proton Energy [MeV]& $\alpha$ [$10^{-17}$ A\,cm$^{-1}$] &$\kappa$\\
\hline
24.3 & 8.8 $\pm$ 0.9 & 2.19 $\pm$ 0.22\\
16.4 & 10.9 $\pm$ 1.1 & 2.73 $\pm$ 0.27\\
10.5 & 12.2 $\pm$ 1.5 & 3.07 $\pm$ 0.37\\
\hline
\end{tabular}
\end{table}

\begin{figure}[htbp]
\centering
\includegraphics[width=.75\textwidth]{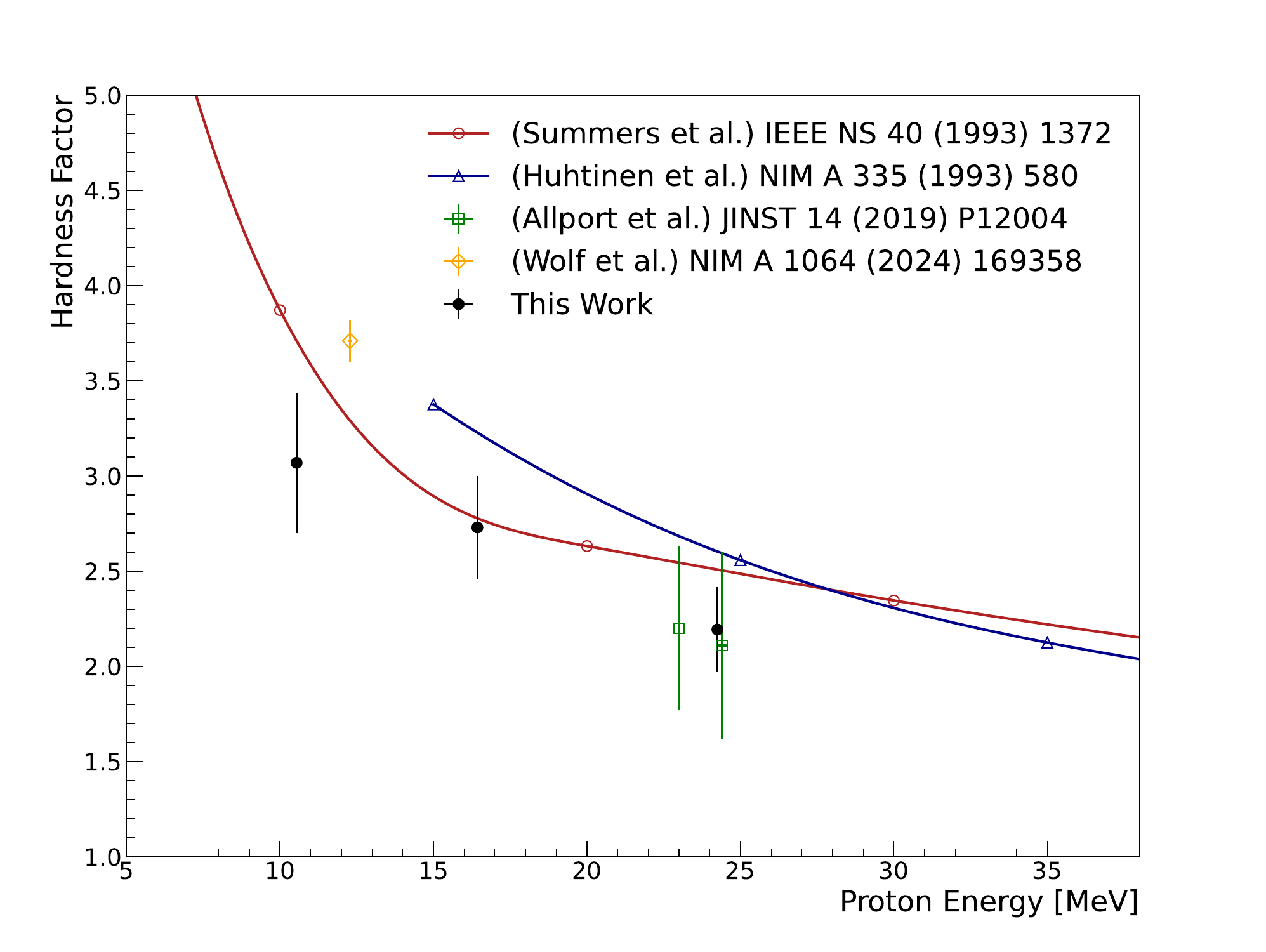}
\caption{The proton hardness factors measured in this work alongside other published results around the same proton energies compared to the tabulated values. The curves for the tabulated values have been interpolated from the discrete predictions.\label{fig:HFSummaryPlot}}
\end{figure}

Figure~\ref{fig:HFSummaryPlot} shows the proton hardness factors measured in this study agaist the tabulated values employed by the RD50 collaboration \cite{RD50HFTables, Summers1993DamageCI, Huhtinen:1993np}. The curves for the tabulated values have been interpolated from the discrete predictions. In the energy range of interest, the hardness factors in the study described by Ref. \cite{Allport:2019kvs} are plotted for samples irradiated at the Birmingham cyclotron and the Karlsruhe cyclotron, at the Irradiation Center Karlsruhe \cite{KarlsruheCyclo}, with proton energies of 24.4 MeV and 23 MeV respectively. Also included is the hardness factor for 12.3 MeV protons from the Bonn Isochronous Cyclotron \cite{Wolf:2024mui}. The hardness factor for 24.3 MeV protons from this study is in good agreement with the previous measurement at the same facility. The hardness factor at 16.4 MeV agrees well with the tabulated values, whereas the 24.3 MeV and 10.5 MeV values suggests a lower bulk damage than predicted. 

The uncertainty in the hardness factor measurements is dominated by the uncertainties associated with the proton fluence calculations. The statistical uncertainty on each fluence measurement is associated with the count rate of the foil during the dosimetry, contributing roughly a 5\% relative uncertainty. The systematic uncertainties are dominated by two main factors, the cross-section of the beam monitor reaction and the efficiency calibration of the HPGe detector. The detector efficiency uncertainty is around 8\%, applying for all data points. The uncertainty from the cross-section of the beam monitor reactions are both around 4\%, though treated as uncorrelated between the 24.3 MeV and 16.4 MeV samples and the 10.5 MeV samples due to the different foils used. There is an additional uncertainty for the 24.3 MeV and 10.5 MeV samples at the lower proton fluences around 10$^{12}$ cm$^{-2}$ due to an alteration to the HPGe detector setup required in order to acquire an adequate count rate. An additional 6\% and 25\% systematic uncertainty component is included in the proton fluence measurements for the 24.3 MeV and 10.5 MeV lower fluence samples, respectively.

The uncertainty in the measured current is dominated by the use of Eq.~\eqref{eqn:CurrentScale} to extrapolate to 20$^\circ$C. To estimate this uncertainty, IV measurements were performed at 20$^\circ$C and the leakage current determined at -500 V bias voltage. Only the lowest fluence samples were used to ensure the leakage currents remained low to avoid significant sensor self-heating. An uncertainty of 10\% was assigned based on the relative difference between the measured and extrapolated leakage current at 20$^\circ$C, when assuming an activation energy of 1.21 eV. Given that the current scaling is also sensitive to the measurement temperature, an ice bath was created to assess the absolute temperature accuracy of the K-Type thermocouple and data logger. The ice bath is expected to provide a reference temperature of 0.0 $\pm$ 0.1$^\circ$C. Any offset in the temperature measurement is corrected for the IV measurements before the applying the current-temperature scaling equation. The statistical uncertainty from repeated IV measurements is negligible. 

Table~\ref{tab:HFCorrelation} summarises the correlation between the hardness factor measurements. The 24.3 MeV and 16.4 MeV measurements are highly correlated due to sharing the same systematic error from the cross-section of the beam monitor reactions for the dosimetry. 

\begin{table}[htbp]
\centering
\caption{Correlation coefficients between the different hardness factor measurements, represented by the proton energy at the sensor.\label{tab:HFCorrelation}}
\smallskip
\begin{tabular}{|c|c|c|}
\hline
\multicolumn{2}{|c|}{Hardness Factor Measurement} & \multirow{2}{*}{Correlation Coefficient, $\rho_{\mathrm{AB}}$}\\
\cline{1-2}
\hspace*{1.2cm} A \hspace*{1.2cm} & \hspace*{1.2cm} B \hspace*{1.2cm} & \\
\cline{1-3}
24.3 MeV & 16.4 MeV & + 0.82\\
24.3 MeV & 10.5 MeV & + 0.48\\
16.4 MeV & 10.5 MeV & + 0.50\\
\cline{1-3}
\end{tabular}
\end{table}

\begin{table}[htbp]
\centering
\caption{The ratios of the current-related damage factors for the three measurements, labelled with the proton energy at the sensor in MeV.\label{tab:AlphaRatios}}
\smallskip
\begin{tabular}{|c|c|}
\hline
$\alpha$ Ratio & Value \\
\hline
$\alpha_{16.4}$/$\alpha_{24.3}$ &  1.24 $\pm$ 0.07 \\
$\alpha_{10.5}$/$\alpha_{24.3}$ &  1.40 $\pm$ 0.16 \\
$\alpha_{10.5}$/$\alpha_{16.4}$ &  1.12 $\pm$ 0.12 \\
\hline
\end{tabular}
\end{table}

Ratios of hardness factors measured at different energies are also presented in Table~\ref{tab:AlphaRatios}. Such ratios are independent of $\alpha_{\text{n}_{\text{eq}}}$ and benefit from cancellations in the impact of some components of the fluence uncertainties, providing a more precise observable for comparison to theoretical predictions.



\section{Conclusions}
\label{sec:conclusion}
The proton hardness factor in silicon was determined for three different proton energies by analysing the change in leakage current of n-in-p FZ pad diodes post-irradiation. The proton energy incident on the samples are 24.3 MeV, 16.4 MeV and 10.5 MeV, leading to hardness factors of $2.19 \pm 0.22$, 2.73 $\pm$ 0.27 and 3.07 $\pm$ 0.37 respectively. The hardness factor measurement at 24.3 MeV agrees well with a previous measurement at the same irradiation facility, but improves the precision by a factor of approximately two. The measurement in this work at 10.5 MeV is notably lower than that of Ref.~\cite{Wolf:2024mui} at 12.3 MeV in a manner which seems inconsistent with the energy dependence expected from Ref.~\cite{Summers1993DamageCI}. The origin of this discrepancy is unclear. One notable difference between the two measurements is the devices used, which differ in technology and bulk thickness. This discrepancy may be clarified with further measurements in this energy region with a variety of different silicon sensor devices. The hardness factors at 24.3 MeV and 10.5 MeV are shown to be lower than the values from theoretical predictions, whereas the hardness factor for 16.4 MeV agrees within its uncertainty. The dominating uncertainty in the hardness factors arises from the fluence calculations, specifically the detector efficiency and the cross-section measurements of the beam monitor reactions.


\acknowledgments
We would like to thank the operations team at the University of Birmingham MC40 cyclotron facility for their help and support, and the ATLAS ITk strip sensor community for access to the devices used in this study. This project has received funding from the European Union's Horizon Europe Research and Innovation programme under Grant Agreement No. 101057511.










\bibliographystyle{JHEP}
\bibliography{biblio.bib}

\end{document}